\documentclass[%
preprint,
 amsmath,amssymb,
 aps,
 ]{revtex4-1}

\usepackage{mathtools,amsmath}
\usepackage{supertabular}
\usepackage{slashed}
\usepackage{amsfonts}
\usepackage{amsmath}
\usepackage{booktabs}
\usepackage{siunitx}
\usepackage{multirow}
\usepackage{colortbl}
\usepackage{xcolor}
\usepackage{graphicx}
\usepackage[utf8]{inputenc}
\usepackage[%
  colorlinks=true,
  urlcolor=blue,
  linkcolor=blue,
  citecolor=blue
]{hyperref}

\begin{document}
\title{Strong coupling constants of negative parity heavy baryons with $\pi$ and $K$ mesons}

\author{T.M.Aliev}
  \email{taliev@metu.edu.tr}
 \author{S. Bilmis}%
   \email{sbilmis@metu.edu.tr}
   \author{M.~Savci}%
   \email{savci@metu.edu.tr}
   \affiliation{%
Department of Physics, Middle East Technical University, 06800, Ankara, Turkey}%

\date{\today}
\begin{abstract}
The strong coupling constants of negative parity heavy baryons belonging to a sextet and antitriplet representations of $SU_f(3)$ with light $\pi$ and $K$ mesons are estimated within the light cone QCD sum rules. It is observed that each class of the sextet-sextet, sextet-antitriplet, and antitriplet-antitriplet transitions can be described by only one corresponding function. The pollutions arising from the positive to positive, positive to negative, and negative to positive parity baryons transitions are eliminated by constructing sum rules for different Lorentz structures. The obtained coupling constants are compared with the ones for the positive parity heavy baryons.
\end{abstract} 
%
\maketitle

\section{Introduction}
\label{sec:intro}
Recent years have been very productive in heavy baryon spectroscopy. The new states with $ J^P = \frac{1}{2}^+$ ($\Lambda_c^{+}$, $\Xi_c^{+}$) and $ J^P= \frac{1}{2}^-$ ($\Lambda_c^{+}(2593)$, $\Xi_c^{+}(2790)$, $\Xi_c^0(2790)$) as well as $J^P = \frac{1}{2}^+$ ($\Omega_c^{(*)}$, $\Sigma_c^{(*)}$) and $ J^P = \frac{3}{2}^+$ ($\Xi_c^{(*)}$)  states have been discovered~\cite{Olive:2016xmw}. Lately, LHCb Collaboration has reported the observation of the five extremely narrow excited $\Omega_c$ baryons decaying into $\Xi_c^{+} K^{-}$~\cite{Aaij:2017nav}. One possible explanation for the existence of these baryons is that they could be the bound states of $P$ wave $ss$-diquark and  $c$-quark~\cite{Karliner:2017kfm}. Up to now, the quantum numbers of newly observed $\Omega_c$ states have not been identified yet. Therefore, the phenomenological studies play a crucial role in this subject. We hope that in the near future more detailed analysis of experiments will be carried out and the quantum numbers will be determined. We stress that the observation of new $\Omega_c$ baryons paves the way for focusing on the studies of the properties of the negative parity baryons. A comprehensive theoretical study of weak, electromagnetic and strong decays require particular attention. The better way to distinguish the particles with different parity from the same final state is studying various decay channels, where one of the main parameters is the strong coupling constant.
hadrons can give useful information about the quark structure of them.

Present work is devoted to the calculation of the strong coupling constants of negative parity heavy baryons with pseudoscalar $\pi$ and $K$ mesons within the light cone QCD sum rules (LCSR) (see for example~\cite{Balitsky:1989ry}). The strong coupling constants are the fundamental quantities to determine the strength of the strong interaction among the participated particles as well as playing an essential role for understanding the structure of hadrons. Note that, the coupling constants of positive parity heavy baryons with $\pi$ and $K$ mesons are calculated in~\cite{Aliev:2010yx}. A comparison study of strong decay constants for positive and negative parity states give insight into underlying QCD dynamics at large distance.

The paper is organized as follows. In Section~\ref{sec:2}, the sum rules for the strong coupling constants of negative parity heavy baryons with pseudoscalar mesons ($BBP$)  are derived. In Section~\ref{sec:numeric}, the numerical analysis of the derived sum rules for BBP couplings is performed. Section~\ref{sec:conclusion} contains our conclusion.

\section{Theoretical Framework}
\label{sec:2}
It is well known that the heavy baryons with single heavy quark belong either to a symmetric $6_F$ or antisymmetric $\bar{3}_F$ flavor representation of $SU(3)_F$ group. From spin, color and flavor symmetry properties it follows that the particles belonging to $6_F$ representation have total spin $1/2$ or $3/2$, while the particles belonging to $\bar{3}_F$ representation only have spin $1/2$.

In order to calculate the $BBP$ coupling constants in the framework of the light cone QCD sum rules, we consider the following correlation function
\begin{equation}
  \label{eq:1}
  \Pi^{(\alpha \rho)} = i \int d^4x e^{iqx} \langle P(q) | \eta^{(\alpha)}(x) \bar{\eta}^{(\rho)}(0) | 0 \rangle 
\end{equation}
where $\alpha = \rho =S$, $\alpha = S, \rho = A$ and $\alpha = \rho = A$ cases correspond to sextet-sextet, sextet-antitriplet, and antitriplet-antitriplet transitions, respectively.
The interpolating currents for the spin-$1/2$ sextet and antitriplet heavy baryons can be written as
\begin{equation}
  \label{eq:2}
  \begin{split}
    \eta^{(S)} = - \frac{1}{\sqrt{2}} \epsilon^{abc} \bigg\{ & (q_1^{aT} C Q^b) \gamma_5 q_2^c +\beta (q_1^{aT} C \gamma_5 Q^b) q_2^c \\ &-\big[ (Q^{aT} C q_2^b) q_1^c + \beta (Q^{aT} C \gamma_5 q_2^b) q_1^c \big] \bigg\}
  \end{split}
\end{equation}

\begin{equation}
  \label{eq:3}
  \begin{split}
    \eta^{(A)} =  \frac{1}{\sqrt{6}} \epsilon^{abc} \bigg\{ & 2 (q_1^{aT} C q_2^b) \gamma_5 Q^c + 2 \beta (q_1^{aT} C \gamma_5 q_2^b) Q^c) q \\ &+ (q_1^{aT} C Q^b) \gamma_5 q_2^c + \beta (q_1^{aT} C \gamma_5 Q^b) q_2^c \\
    &+  (Q^{aT} C q_2^b) \gamma_5 q_1^c + \beta (Q^{aT} C \gamma_5 q_2^b) q_1^c  \bigg\}    
  \end{split}
\end{equation}
where $a,b,c$ are the color indices and $\beta$ is the arbitrary parameter, and $\beta = -1$ case corresponds to Ioffe current which we shall use in further discussions. The light quark content of sextet and antitriplet heavy baryons are given in Table~\ref{tab:1}.

\begin{table*}[hbt]
  \centering
  \renewcommand{\arraystretch}{1.3}
  \setlength{\tabcolsep}{8pt}
  \begin{supertabular}{lcccccccccc}
    \toprule
    & $\Sigma_{b(c)}^{+(++)}$ & $\Sigma_{b(c)}^{0(+)}$ & $\Sigma_{b(c)}^{-(0)}$ & $\Xi_{b(c)}^{-(0)'}$ & $\Xi_{b(c)}^{0(+)'}$  & $\Omega_{b(c)}^{-(0)}$ & $\Lambda_{b(c)}^{0(+)}$ & $\Xi_{b(c)}^{-(0)}$ & $\Xi_{b(c)}^{0(+)}$ \\
    \midrule
$q_1$ & $u$ & $u$ & $d$ &$d$ &$u$ & $s$ & $u$ & $d$ & $u$ \\
    $q_2$ & $u$ & $d$ & $d$ & $s$ & $s$ &$s$ &$d$ &$s$ &$s$ \\
    \bottomrule
  \end{supertabular}
  \caption{The quark flavors $q_1$ and $q_2$ for the baryons in the sextet and
the antitriplet representations}
  \label{tab:1}
\end{table*}

The first stage of the calculation of $\Pi^{(\alpha \rho)}$ within the light-cone sum rules is writing it in terms of hadronic parameters. Saturating eq.~(\ref{eq:1}) by corresponding baryons we get,

\begin{equation}
  \label{eq:4}
  \begin{split}
    \Pi^{(\alpha \rho)} = \frac{ \langle 0 | \eta^{(\alpha)}|B_f(p) \rangle \langle B_f(p) P(q)| B_i(p+q) \rangle \langle B_i(p+q)|\bar{\eta}^{(\rho)} |0 \rangle}{ \big( p^2 - m_f^2) \big((p+q)^2 - m_i^2 \big)} + ...
  \end{split}
\end{equation}
where $p$ ($p+q$) and $m_f$ ($m_i$) is the final (initial) state baryon four momentum and mass, respectively.

The interpolating current $\eta^{(S)}$ or $\eta^{(A)}$ interact with both positive and negative parity baryons. Under this circumstance the matrix elements in eq.~(\ref{eq:2}) are defined as
\begin{equation}
  \label{eq:5}
  \begin{split}
    \langle 0 | \eta^{(\alpha)} | B_2^{(+)}(p) \rangle &= \lambda_{+}^{(\alpha)} \bar{u}_2(p) \\
    \langle 0 | \eta^{(\alpha)} | B_2^{(-)}(p) \rangle &= \lambda_{-}^{(\alpha)} \gamma_5 u_2(p) \\
    \langle B_2^{\pm} P(q) | B_i^{\pm} \rangle &= g_{\pm \pm} \bar{u}_2(p) \gamma_5 u_1(p+q) \\
    \langle B_2^{\pm} P(q) | B_i^{\mp} \rangle &= g_{\mp (\pm)} \bar{u}_2(p) i u_1(p+q) \\
  \end{split}
\end{equation}
where $\lambda_{{+}(-)}$ is the residue of the positive (negative) parity baryon and $g_{ij}$ is the coupling constant of the pseudoscalar meson with heavy baryon.

Now let us first consider the sextet-sextet transition. Using eqs.~(\ref{eq:4}) and (\ref{eq:5}) and performing summations over spins of initial and final baryons, we get,
\begin{equation}
  \label{eq:6}
  \begin{split}
    \Pi^{(\alpha \rho)} &=  A_{++}^{(\alpha \rho)} (\slashed{p} + m_{2+}) \gamma_5 (\slashed{p} + \slashed{q} + m_{1+}) \\
    & + A_{--}^{(\alpha \rho)} \gamma_5 (\slashed{p} + m_{2-}) \gamma_5 (\slashed{p} + \slashed{q} + m_{1-})(-\gamma_5) \\
    & + A_{-+}^{(\alpha \rho)} (\slashed{p} + m_{2+}) (\slashed{p} + \slashed{q} + m_{1-})(-\gamma_5) \\
    & + A_{+-}^{(\alpha \rho)} \gamma_5 (\slashed{p} + m_{2-}) (\slashed{p} + \slashed{q} + m_{1+})
 \end{split}
\end{equation}
in which
\begin{equation}
  \label{eq:7}
  \begin{split}
    A_{ij}^{(\alpha \rho)} = \frac{g_{ij}^{(\alpha \rho)} \lambda_{2j}^{(\alpha)} \lambda_{1i}^{(\rho)}}{(m_{2j}^2 - p^2) \big(m_{1i}^2 - (p+q)^2 \big)}
  \end{split}
\end{equation}
and $i(j) = \pm (\pm)$.  

To construct the sum rules for the strong coupling constants of $SSP$, $SAP$, and $AAP$ ($S$, $A$ and $P$ correspond to sextet, antitriplet states, and pseudoscalar meson respectively), the expressions of the correlation functions from the QCD side are needed. Before calculating the $SSP$, $SAP$ and $AAP$ coupling constants, we would like to make the following remark. Considering the approach presented in \cite{Aliev:2006xr}, one can easily show that each class of transitions can be described by only one invariant function. The main advantage of this approach is that relations among the invariant functions become structure independent although the expression of the invariant function is structure dependent. (For more details see \cite{Aliev:2010yx}.) Hence, to determine all the relevant coupling constants, it is enough to consider the $\Sigma_Q^0 \rightarrow \Sigma_Q^0 P$, $\Xi_Q^{\prime 0} \rightarrow \Xi_Q P$ and $\Xi_Q^0 \rightarrow \Xi_Q^{0} P$ transitions. All the other possible strong coupling constants can be obtained from these couplings, i.e. each class of transitions is described only by one coupling constant.

Relations among the invariant functions can easily be obtained by using the approach given in~\cite{Aliev:2006xr}.  It follows from eq.~(\ref{eq:6}) that the correlation function contains contributions from positive-negative, negative-positive and negative-negative transitions. Among them, we need only one, namely, $A_{--}$ which described the transition between negative parity heavy baryons. For this goal, we need to solve four equations obtained from four different Lorentz structures, $\slashed{p} \slashed{q} \gamma_5$, $\slashed{p} \gamma_5$, $\slashed{q} \gamma_5$, and $\gamma_5$.

Now, let turn our attention to the calculation of the correlation function from QCD side. As we have already noted, to construct the sum rules for strong coupling constants for $SSP$, $SAP$, $AAP$ transitions, calculations of the correlation functions from QCD side are required. This can be calculated by using the operator product expansion (OPE) over the twist of nonlocal operators in the deep Euclidean region, where $-p^2 \rightarrow \infty$ and $-(p+q)^2 \rightarrow \infty$.

The QCD part of the correlation function can be obtained by contracting quark fields. In this case, two different contributions appear; perturbative and non-perturbative. The perturbative contribution can be obtained by inserting the free quark propagators to the expression of correlation functions. The expressions of the light and heavy quark propagators are;
\begin{equation}
  \label{eq:8}
  \begin{split}
    S^{ab}(x) &= \frac{i \slashed{x} \delta^{ab}}{2 \pi^2 x^4} - \frac{m_q \delta^{ab}}{4 \pi^2 x^2} 
    - \frac{i}{32 \pi^2} g_s  G_{\mu \nu}^{ab} (\sigma^{\mu \nu} \slashed{x}
    + \slashed{x} \sigma^{\mu \nu}) \\
    &- \frac{m_s}{32 \pi^2}g_s  G_{\mu \nu}^{n} \frac{{(\lambda^{n}})^{ab}}{2} \ln (- \frac{x^2 \Lambda^2}{4} ) \sigma^{\mu \nu}  
      - \frac{m_q \delta^{ab}}{2^9 3 \pi^2} x^2 \ln(-\frac{x^2 \Lambda^2}{4}) \langle g_s^2 G^2 \rangle + ...
  \end{split}
\end{equation}
and
\begin{equation}
  \label{eq:9}
  \begin{split}
    S_Q^{ab}(x) &= \frac{m_Q^3 \delta^{ab}}{2 \pi^2} \big\{ \frac{m_Q i \slashed{x}}{m_Q^2 (\sqrt{-x^2})^2} K_2 (m_Q \sqrt{-x^2}) + \frac{1}{m_Q \sqrt{-x^2}} K_1 (m_Q \sqrt{-x^2}) \big\} \\
    &- \frac{m_Q g_s G_{\mu\nu}^{ab}}{8 (2 \pi)^2} \big\{m_Q i (\sigma^{\mu \nu} \slashed{x} + \slashed{x} \sigma^{\mu \nu}) \frac{K_1 (m_Q\sqrt{-x^2})}{m_Q \sqrt{-x^2}} + 2 \sigma^{\mu\nu} K_0(m_Q \sqrt{-x^2}) \big\}
  \end{split}
\end{equation}
where $K_i$ are the modified Bessel functions of the second kind, $m_q$ and $m_Q$ are the light and heavy quark mass respectively. $\Lambda$ is the parameter for separating the perturbative and non-perturbative regions and for numerical calculations we use $\Lambda= 0.5~\rm{GeV}$~\cite{Chetyrkin:2007vm}. On the other hand, the non-perturbative contribution can be obtained by replacing one of the light quark propagators by

\begin{equation}
  \label{eq:14}
S_{\alpha \beta}^{ab} \rightarrow  -\frac{1}{4} \Gamma_{\beta \alpha}^i (\bar{q}^a \Gamma^i q^b)
\end{equation}
where $\Gamma^i$ is the full set of Dirac matrices. Then this expression is sandwiched between the pseudoscalar meson and vacuum states. These matrix elements as well as the ones $\langle P(q) | \bar{q} \Gamma G_{\mu \nu} q | 0 \rangle$, where $G_{\mu \nu}$ is the gluon field strength tensor, are determined in terms of pseudoscalar meson distribution amplitudes (DA). Up to twist-4 the explicit expressions of the pseudoscalar meson DA's are presented in \cite{Ball:1998je,Ball:2006wn}. In the present study we neglect the four-particle contributions due to their small values (see for example \cite{Aliev:2010su}.) 
Using these expressions for the propagators and DA's for the pseudoscalar mesons, the correlation functions from the QCD side can be calculated. 

Equating the coefficients of the Lorentz structures $\slashed{p} \slashed{q} \gamma_5$, $\slashed{p} \gamma_5 $, $\slashed{q} \gamma_5$, and $\gamma_5$ in hadronic and QCD sides and performing the Borel transformations over $p^2$ and $(p+q)^2$ in order to suppress the contributions of higher states and continuum, we get the following four linear equations:

\begin{equation}
  \label{eq:10}
  \begin{split}
  \Pi_1^{(\alpha \rho)B} &=  -A_{++}^{(\alpha \rho)B} + A_{--}^{(\alpha \rho)B} -A_{-+}^{(\alpha \rho)B} + A_{+-}^{(\alpha \rho)B}  \\
  \Pi_2^{(\alpha \rho)B} &=  (m_{1+} - m_{2+}) A_{++}^{(\alpha \rho)B} + A_{--}^{(\alpha \rho)B} (m_{1-} - m_{2+}) - (m_{2+} + m_{1-}) A_{-+}^{(\alpha \rho)B} - (m_{2-} + m_{1+}) A_{+-}^{(\alpha \rho)B} \\
  \Pi_3^{(\alpha \rho) B} & = -m_{2+} A_{++}^{(\alpha \rho)B} - m_{2-} A_{--}^{(\alpha \rho)B} - m_{2+} A_{-+}^{(\alpha \rho)B} - m_{2-} A_{+-}^{(\alpha \rho)B} \\
  \Pi_4^{(\alpha \rho)B} & = (-m_{1+}^2 + m_{1+} m_{2+}) A_{++}^{(\alpha \rho)B} + (-m_{2-} m_{1-} + m_{1-}^2) A_{--}^{(\alpha \rho)B} - (m_{2+} m_{1-} + m_{1-}^2) A_{-+}^{(\alpha \rho)B} \\
  &+ (m_{2-} m_{1+} + m_{1+}^2) A_{+-}^{(\alpha \rho)B}
\end{split}
\end{equation}

where $A_{ij}^{(\alpha \rho) B} = g_{ij}^{(\alpha \rho)} \lambda_{1i}^{(\alpha)} \lambda_{2j}^{(\rho)} e^{-\frac{m_{1i}^2}{M_1^2} - \frac{m_{2j}^2}{M_2^2}}$. 

Solving eq.(\ref{eq:10}) for $A_{--}^{(\alpha \rho) B}$ we get the desired sum rules for the strong coupling constant $g_{--}$, responsible for the $SSP$, $SAP$ and $AAP$ transitions.

\begin{equation}
  \label{eq:15}
  \begin{split}
    g_{--}^{(\alpha \rho)} &= \frac{ e^{\big[\frac{m_{1^-}^2}{M_1^2} + \frac{m_{2^-}^2}{M_2^2}\big]} }{\big[ (2 m_{1-} m_{2-}+m_{2+} (2 m_{1+} + m_{2-} - m_{2+})) \big] \lambda_{1^-}^{(\alpha)} \lambda_{2^-}^{\rho}} 
     \times \bigg(  m_{1-} m_{1+} \Pi_1^{(\alpha \rho) B}  \\ &- \Pi_3^{(\alpha \rho) B} (m_{1-} + m_{2+})  + m_{1-} \Pi_2^{(\alpha \rho) B} + m_{1+} m_{2+} \Pi_1^{(\alpha \rho) B} - m_{1+} \Pi_2^{(\alpha \rho) B} + m_{2+} \Pi_2^{(\alpha \rho) B} -\Pi_4^{(\alpha \rho) B} \bigg)
  \end{split}
\end{equation}

The expressions of $\Pi_i^{(\alpha \rho) B}$ responsible for $SSP$, $SAP$, and $AAP$ are very lengthy, and for this reason, we do not present all of them here.

In eq.~(\ref{eq:15}), $M_1^2$ and $M_2^2$ are the Borel mass square parameters. Due to the fact that initial and final baryon masses are very close to each other, we take $M_1^2 = M_2^2 = 2M^2$. The residues of $\lambda_{1i}^{(\alpha)}$ and $\lambda_{2j}^{(\rho)}$ of heavy baryons are calculated in~\cite{Aliev:2015qea}.
The continuum subtraction procedure is carried out at $M_1^2 = M_2^2 = 2M^2$ and at $u_0 = \frac{M_1^2}{M_1^2 + M_2^2} = \frac{1}{2}$ point by using the expression
\begin{equation}
  \label{eq:11}
  (M^2)^n e^{-m^2/M^2} \rightarrow \frac{1}{\Gamma(n)} \int_{m_Q^2}^{s_0} ds~e^{-s/M^2} (s-m_Q^2)^{n-1}.
\end{equation}

\section{Numerical Analysis}
\label{sec:numeric}
In the present section, we perform the numerical analysis of the light-cone sum rules for the strong coupling constants of spin $1/2$ negative parity heavy baryons with $\pi$ and $K$ mesons obtained in the previous section. The main input parameters of the LCSR are the pseudoscalar meson DA's which are derived in \cite{Ball:1998je,Ball:2006wn}. The values of other input parameters entering the sum rules are the lepton decay constants $f_\pi$ an $f_K$ whose values are $f_\pi = 0.13~\rm{GeV}$ and $f_K=0.16~\rm{GeV}$. For performing numerical analysis, the mass of the $u$ and $d$ quarks are taken as zero and for the mass of strange, charm and beauty quarks we used: $m_s = (96^{+8}_{-4})~\rm{MeV}$, $m_c = (1.27 \pm 0.03)~\rm{GeV}$ and $m_b = (4.18^{+0.04}_{-0.03})~\rm{GeV}$, and we put renormalization scale $\mu =1~\rm{GeV}$. If the renormalization scale is chosen as $\mu = 2~\rm{GeV}$, the results change $(5-8)\%$ at most.

The sum rules for the coupling constants describing $SSP$, $SAP$, and $AAP$ transitions contain two auxiliary parameters (when $\beta = -1$), continuum threshold $s_0$ and Borel mass parameter $M^2$. One of the main issues of the sum rules is to find the so-called working region of these parameters where the coupling constants are insensitive to their variation. The limits of the $M^2$ is obtained from the conditions that, the pole contributions should be dominant over the higher states and continuum contributions as well as the convergence of operator product expansion. Our numerical analysis shows that these conditions are satisfied when $M^2$ change in the regions:

\begin{table}[hbt]
  \renewcommand{\arraystretch}{1.3}
  \setlength{\tabcolsep}{8pt}
      \begin{tabular}{ccr}
        \multirow{2}{*}{Channel} & \multicolumn{2}{c}{Strong Coupling Constant}  \\
        \multicolumn{1}{c}{} & Negative Parity & ~~~Positive Parity\\
        \midrule
        $\Xi_b^{\prime-} \rightarrow \Xi_b^- K^+$ & $3 \pm 1$ & $9.8 \pm 3.5$ \\
       $ \Omega_b^{-} \rightarrow \Xi_b^{-} \bar{K}^0$ & $5 \pm 1$ & $13.5 \pm 4.8$\\
       $ \Sigma_b^0 \rightarrow \Xi_b^0 \bar{K}^0$ & $5 \pm 1$ & $8.9 \pm 3.1$ \\
       $ \Xi_b^{\prime 0} \rightarrow \Sigma_b^{+} K^-$ & $9 \pm 1$ & $10.0 \pm 3.6$ \\
       $ \Omega_b^{-} \rightarrow \Xi_b^{\prime 0} K^-$ & $10 \pm 2$ & $12.3 \pm 4.4$ \\ 
       $ \Sigma_b^{-} \rightarrow $ $\Lambda_b^0 \pi^-$ & $8 \pm 2$ & $11.5 \pm 3.9$ \\
       $ \Xi_b^{\prime 0} \rightarrow \Xi_b^0 \pi^0$ & $4 \pm 1$ & $6.1 \pm 2.2$ \\
       $ \Sigma_b^0 \rightarrow \Sigma_b^- \pi^+$ & $8 \pm 2$ & $13 \pm 4.5$ \\
       $ \Xi_b^{\prime 0} \rightarrow \Xi_b^{\prime 0} \pi^0$ & $4 \pm 1$ & $7.3 \pm 2.6$ \\
      \bottomrule
      \end{tabular}
  \caption{}
  \label{tab:2}
\end{table}

\begin{table}[hbt]
  \renewcommand{\arraystretch}{1.3}
  \setlength{\tabcolsep}{8pt}
      \begin{tabular}{ccr}
        \multirow{2}{*}{Channel} & \multicolumn{2}{c}{Strong Coupling Constant}  \\
        \multicolumn{1}{c}{} & Negative Parity & ~~~Positive Parity\\
        \midrule
        $\Xi_c^{\prime 0} \rightarrow \Xi_c^0 K^+$ & $1.5 \pm 0.5$ & $ 2.1 \pm 0.8$ \\
        $ \Omega_c^0 \rightarrow \Xi_c^0 \bar{K}^0$ &  $2.5 \pm 0.5$ & $ 3.0 \pm 1.1$ \\
        $ \Sigma_c^{+} \rightarrow \Xi_c^{+} K^0$ & $2 \pm 1$ & $ 3.7 \pm 1.3$ \\
        $ \Xi_c^{\prime +} \rightarrow \Xi_c^{++} K^-$ & $10 \pm 2$ & $ 9.0 \pm 1.0$ \\
        $ \Omega_c^0 \rightarrow \Xi^{\prime +} K^0$ & $11 \pm 1$ & $ 5.6 \pm 1.9$ \\
        $ \Sigma_c^0 \rightarrow \Lambda_c^{+} \pi^-$ & $3 \pm 1$ & $ 5.6 \pm 1.8$ \\
        $ \Xi_c^{\prime +} \rightarrow \Xi_c \pi^0$ & $1.5 \pm 0.5$ & $ 2.0 \pm 0.7$ \\
        $ \Sigma_c^{+} \rightarrow \Sigma_c^{0} \pi^+$ & $11 \pm 2$ &  $4.1 \pm 1.5$ \\
        $ \Xi_c^{\prime +} \rightarrow \Xi_c^{\prime +} \pi^0$ & $5 \pm 1$ & $ 3.0 \pm 1.1 $ \\
        \bottomrule
      \end{tabular}
  \caption{}
  \label{tab:3}
\end{table}
\begin{itemize}
\item For b-sector:
  \begin{equation}
    \label{eq:13}
    \begin{split}
      &5~\rm{GeV^2} \leq M^2 \leq 10~\rm{GeV^2} \hspace{2cm} (\text{for}~~\Lambda_b, \Sigma_b)\\
      &5~\rm{GeV^2} \leq M^2 \leq 11~\rm{GeV^2} \hspace{2cm} (\text{for}~~\Xi_b, \Xi_b^\prime)\\
      &6~\rm{GeV^2} \leq M^2 \leq 12~\rm{GeV^2} \hspace{2cm} (\text{for}~~\Omega_b)
    \end{split}
  \end{equation}
\item For c-sector:
  \begin{equation}
    \label{eq:13}
    \begin{split}
      &3~\rm{GeV^2} \leq M^2 \leq 6~\rm{GeV^2} \hspace{2cm} (\text{for}~~\Lambda_c, \Sigma_c)\\
      &3~\rm{GeV^2} \leq M^2 \leq 7~\rm{GeV^2} \hspace{2cm} (\text{for}~~\Xi_c, \Xi_c^\prime)\\
      &4~\rm{GeV^2} \leq M^2 \leq 8~\rm{GeV^2} \hspace{2cm} (\text{for}~~\Omega_c)
    \end{split}
  \end{equation}
\end{itemize}

\begin{figure}[hb]
  \centering
  \includegraphics[scale=0.70]{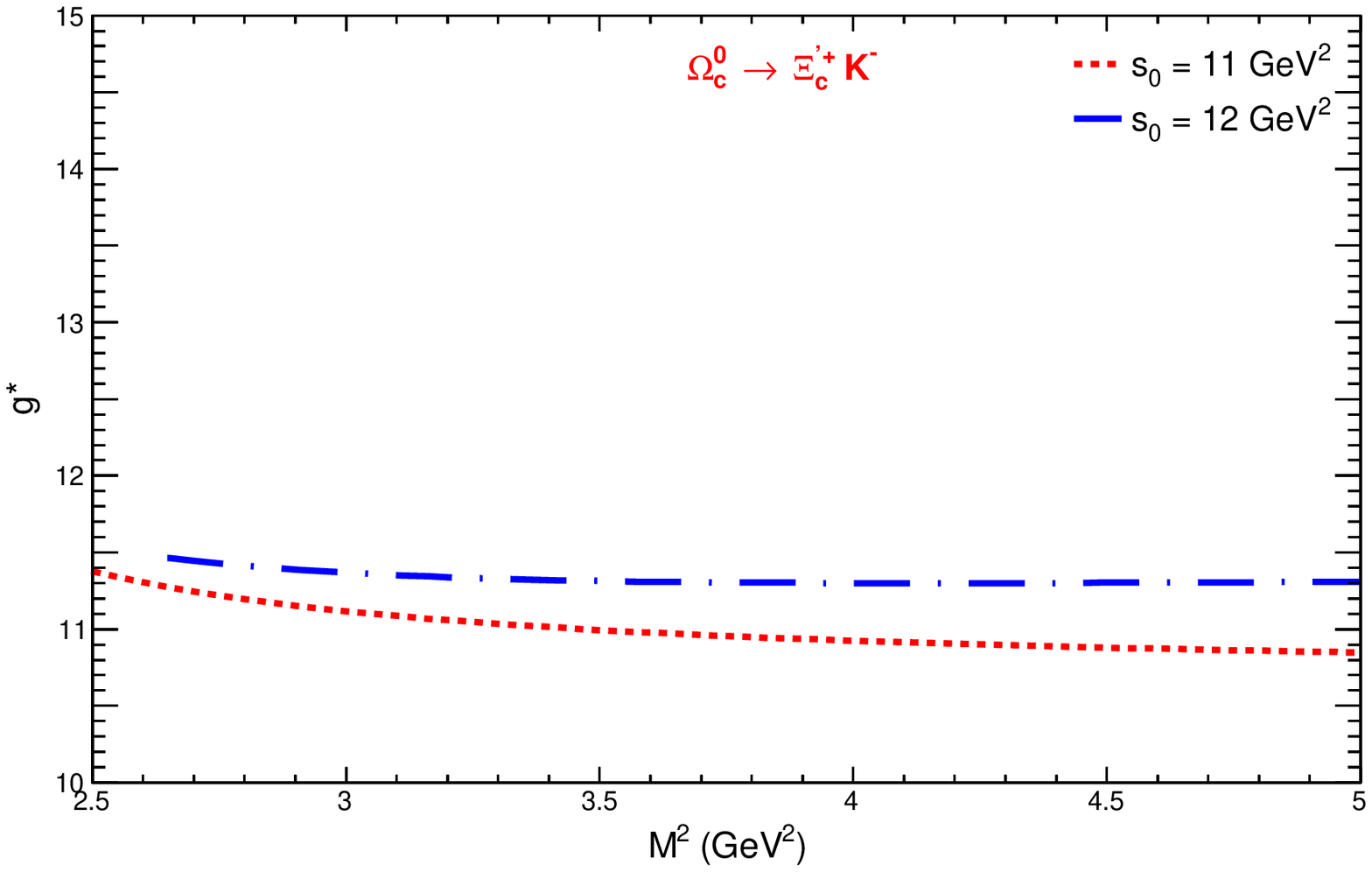}
  \caption{The dependence of the strong coupling constant for $\Omega_c \rightarrow \Xi_c^{\prime +} K$ transition on $M^2$ at two fixed values of $s_0$.}
  \label{fig:1}
\end{figure}

\begin{figure}[hb]
  \centering
  \includegraphics[scale=0.70]{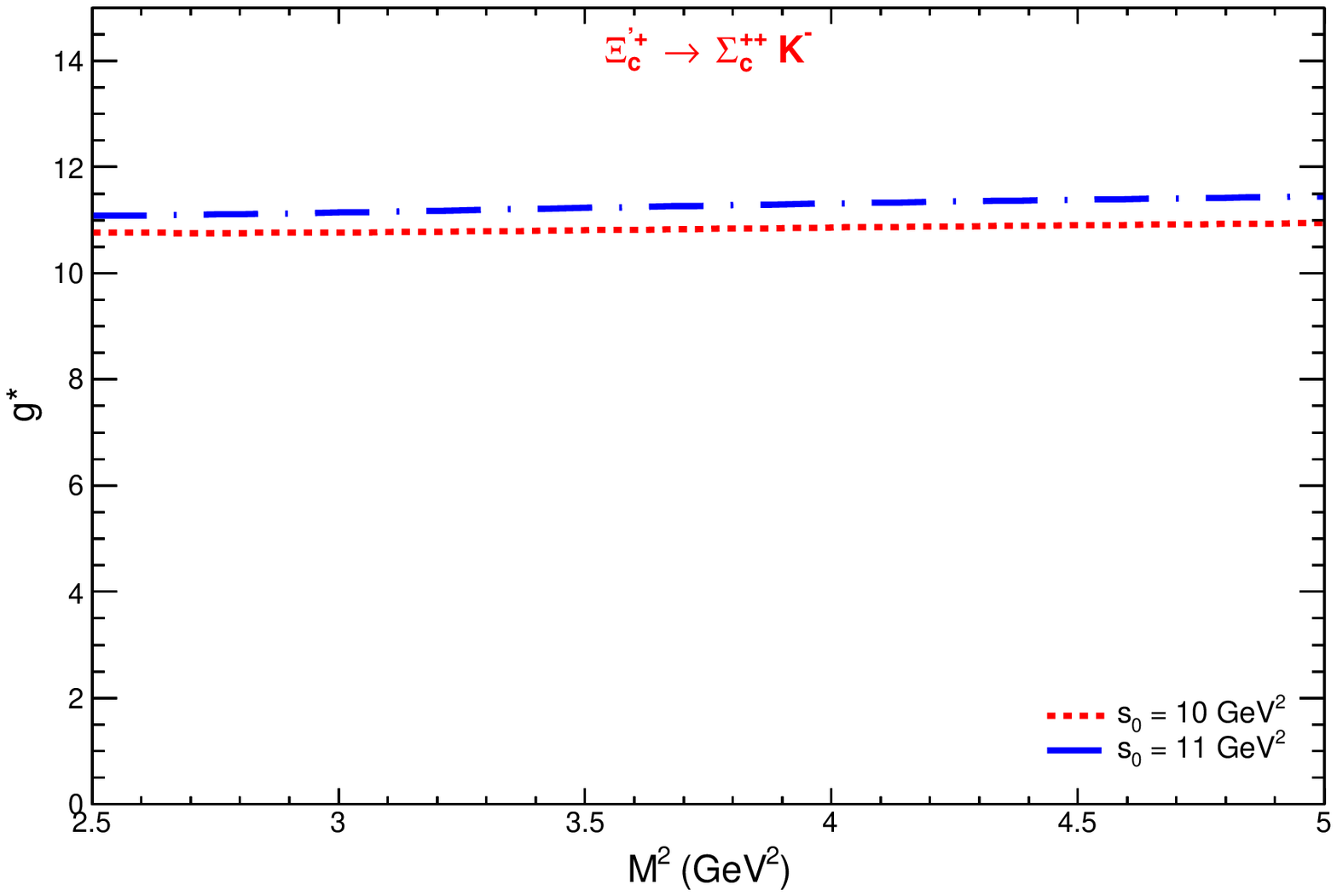}
  \caption{The same as in Fig.~(\ref{fig:1}), but for $\Xi_c^{\prime +} \rightarrow \Sigma_c^{++} K^-$ transition.}
  \label{fig:2}
\end{figure}

\begin{figure}
  \centering
  \includegraphics[scale=0.70]{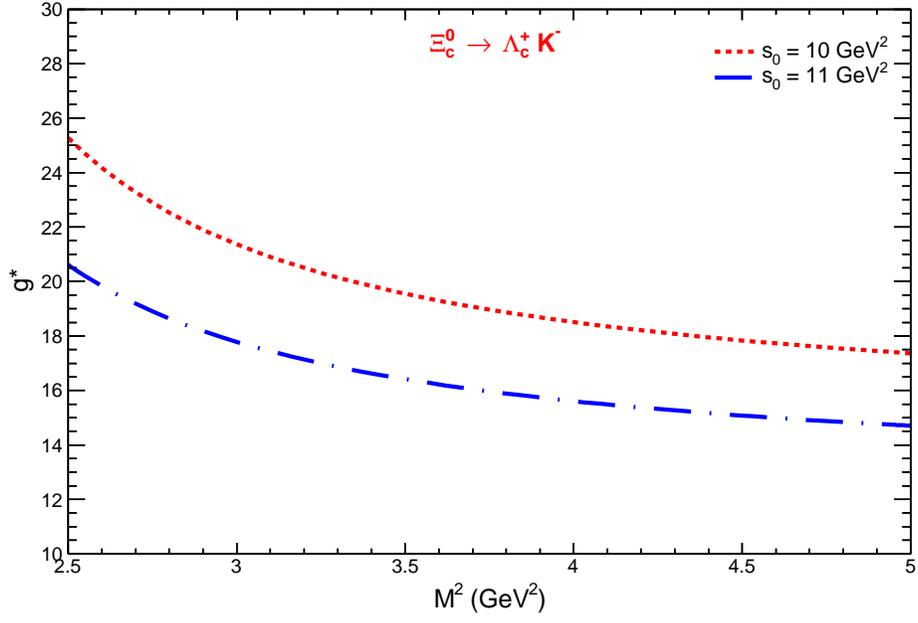}
  \caption{The same as in Fig.~(\ref{fig:1}), but for $\Xi_c^{0} \rightarrow \Lambda_c^+ K^-$ transition.}
  \label{fig:6}
\end{figure}

The continuum threshold in sum rules is related with the first excited energy state, $\sqrt{s_0} - m_{\text{ground}} = \Delta$. Analysis of the various sum rules leads to the result that usually $\Delta$ varies between $0.3$ and $0.8$~\rm{GeV}. In our numerical analysis for $\Delta$, we used $\Delta = 0.4; 0.5~\rm{GeV}$. As an example, in Figs.(\ref{fig:1}), (\ref{fig:2}), and (\ref{fig:6}), we present the dependence of the coupling constant for $\Omega_c^0 \rightarrow \Xi_c^{\prime 0} K$, $\Xi_c^{\prime +} \rightarrow \Sigma_c^{++} K^-$ and $\Xi_c^0 \rightarrow \Lambda_c^{+} K^-$ transitions on $M^2$ at two fixed values of $s_0$. From these figures we observe that $g_{--}$ exhibits good stability with respect to the variation of $M^2$ in the ``working region''. The results are weakly dependent on the $s_0$ values and we get,

\begin{equation}
  \label{eq:16}
  g_{--} =\begin{cases}
    (10 \pm 2), & \text{for $(\Omega_b \rightarrow \Xi_b^{\prime 0} K^-$)}.\\
    (10 \pm 2), & \text{for $ (\Xi_c^{\prime +} \rightarrow \Xi_c^{++} K^-$}) \\
    (10.5 \pm 1.0), & \text{for $ (\Xi_c^0 \rightarrow \Lambda_c^{+} K^-$)} 
  \end{cases}
\end{equation}
A similar analysis for strong coupling constants of the other channels are performed, and the obtained values are depicted in Table~\ref{tab:2} and \ref{tab:3}. In this table, we also present the values of the strong coupling constants of the corresponding positive parity baryons for completeness.

The errors in the values of the $g_{--}$ included uncertainties coming from the variation of the $s_0$, $M^2$ and the values of the input parameters.
Comparing these results, we see that the strong coupling constants of heavy hadrons with b-quark with pseudoscalar mesons for positive parity baryons are larger than similar couplings for negative parity ones. In the transitions between heavy hadrons with c-quark, the situation is analogous to the previous case. Only for a few number of transitions, the coupling constants of the negative parity heavy baryons with pseudoscalar mesons exceeds the similar couplings for positive parity ones.

Our final remark is about the form of the interpolating current. As we already noted that the interpolating current contains auxiliary parameters $\beta$. We perform calculations at $\beta = -1$ (Ioffe current). In order to compare the results for strong coupling constants when general current and Ioffe current used, the results of this present work should be recalculated, and the numerical analysis needs to be reperformed.
\section{Conclusion}
\label{sec:conclusion}
In this work, we estimate the strong coupling constants of the negative parity heavy baryons with pseudoscalar mesons within the light cone QCD sum rules. It is emphasized that each class of transitions, namely sextet-sextet, sextet-antitriplet and antitriplet-antitriplet transitions can be described by only one corresponding invariant function. The values of strong coupling constants for the transitions under consideration are presented, and the results are compared with the similar couplings for positive parity heavy baryons.
\section{Acknowledgements}
The authors thank METU-BAP Grant No. 01-05-2017-004. 
\bibliography{strong}

 \clearpage

\end{document}